# Real and simulated CBM data interacting with an ESCAPE datalake

*E. Clerkin*[*,1], *P.-N. Kramp*[2], *P.-A. Loizeau*[1], *and M. Szuba*[2]

[1]FAIR, Darmstadt, Germany; [2]GSI, Darmstadt, Germany

## Introduction

ESCAPE is an on-going EU-funded Horizon 2020 project whose aim is to standardise, store and share data from large scale European astronomical and particle physics infrastructure under the auspicious of open science, sharing data, and possibility of leveraging efforts of citizen scientists worldwide towards astronomy and experimental particle physics research. The FAIR particle accelerator in Darmstadt is a member of this collaborative cluster and the CBM experiment as a pillar of FAIR has been proactive in engaging with testing, and validating the tools for this open science project. In particular, in the case of the CBM experiment, here we document the CBM interacting with the FAIR and CERN datalakes, which are distributed repositories of raw or blobs of semi-structured data used for managing big data, as maintained by other work-packages of the ESCAPE project. For information on ESCAPE, Refs. [1, 2, 3] and the references therein may be consulted.

We wanted to showcase a spectrum of CBM data interacting with ESCAPE data environment in a broad sense. Fig. 1 shows a highly simplified block diagram of possible interaction points of CBM and ESCAPE. The CBM experiment is in its production stage and its software suite CBM-ROOT is still under active development, but we nevertheless wanted to showcase ESCAPE interacting with both experimental and simulated data. On the simulation side, many of the existing tools may be tuned to interact with the ESCAPE datalakes. In particular we will interact with the datalake as our storage medium for input and output of data for transport and digitisation stages. For real experimental data, CBM has a mini CBM project (mCBM; cp. Ref.[4]) which takes beam from the existing GSI SIS18 accelerator and is used by us in the CBM collaboration to test hardware and software of the full CBM. It will here give us the exciting opportunity to showcase ESCAPE interacting directly with a functioning particle physics experiment under real-time data injection conditions. This was the first-time a showcase of this kind was shown. Lastly, to complete the data demonstration, the final task was to interact at the reconstruction of tracks stage which may be thought of as the basic data analysis step. We had the two options to either interact with simulated data or with the mCBM real experimental data. It was decided to use real data from mCBM July 2020 data processing run as a stronger statement of ESCAPE usability. These three demonstrations were prepared during the course of 2021 and the efforts accumulated and showcased during a single week period for the data acquisition challenge (DAC21) in November 2021. In each of the three sections that follow, we report on the success of each of these challenges. This report was compiled from the DAC21 activities logbook. All times are approximate and in UTC.

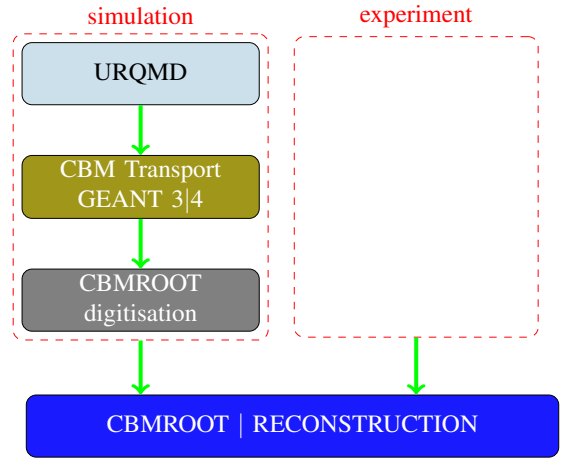

Figure 1: Basic block diagram layout displaying interaction points between the CBM toolkit and external software packages for the CBM experiment at GSI/FAIR.

## mCBM data ingestion

In order to showcase the injection of real-time data, it had been hoped that mCBM cosmic-ray data taking which was originally planned for early November would be available. During late November, this had been rescheduled for December. During the DAC, it was instead decided to replay acquisition of data taken by mCBM in July 2021. From the data-lake point of view, this is functionally the same as if we had data coming directly from the detector. Fig. 2 shows a photo of the mCBM experiment at this time, with the detector acronyms overlaid.

The task and our team was split between a replay side and an injection side. The physical transfer of data was not needed in this demonstration. It was noted that the FAIR datalake was hosted on the same lustre filesystem as was replayed to by the replay team, but instead replicas of the data were registered to the datalake providing its functionality.

---

[*]e.clerkin@gsi.de

*Replay Side*

- 2021-11-23, 15:00 - preparatory scripts for replay start. Several issues needed to be overcome whereby locked-in commands tried to specific ssh-keys were disabled by IT administrators. Our original plan therefore needed a work around and an alternative approach was pursued. The process was achieved by rsyncing of the experimental data over Infiniband at 150-200 MB/s, with 10 parallel copy jobs run on the GSI Green IT Cube cluster, leading to a 2 GB/s average rate. This rate was chosen to mimic the approximate high rate of mCBM data taking. The job array started at 18:32, most of them (9/10) finished around 18:36. One of the files had a much reduced writing speed of 800 kB/s due to an unknown reason. Writing for the final delayed file closed at 19:56.

*Infrastructure used:* As a source, two compute nodes of the miniFles cluster (experiment side cluster of the mCBM), with each 5 HDD (4 TB) holding the original data files of the mCBM 2021 beam campaign. The files were written in parallel so at least one file from each HDD is needed to reconstruct a segment of any mCBM run.

mFLES to Virgo (GSI batch cluster hosted in the Green-Cube building) Infiniband backbone comprises of 10 Virgo logical nodes (SLURM jobs with non-default resources options), with 8 GB RAM each, each executing an rsync process to one of the mFLES HDD for the first 10 files of a typical mCBM run.

*Ingestion Side*

- 2021-11-23, 15:30 - initial setup

- 2021-11-23, 17:30 - status check shows no replicas or rules having been registered with Rucio in spite of the first batch of files having already appeared in the source directory. Problem tracked down to file-system notifications not functioning as expected, began converting the script to polling mode

- 2021-11-23, 18:00 - tests of converted script repeatedly fail on connections to the CERN Rucio server being refused. Eventually discovered that recent changes to GSI network which allowed direct access to FAIR-ROOT from our compute cluster, now require rucio-clients traffic from FAIR-ROOT to the Rucio server to go through a local HTTPS proxy

- 2021-11-23, 18:30 - refactored script launched successfully, with forced delay of 60 seconds between files to avoid pile-up. Replicas begin to be added to the data lake

- 2021-11-24, 13:00 - status check shows 92 replicas registered successfully but owing to a typo, the script only ran once instead of periodically. Ran the script again to register the remaining 8 files

- 2021-11-24, 13:15 - final status check shows all 100 replicas registered successfully. Conducted a random sampling of associated DIDs, all files accessible

*Conclusions*

Achieved asynchronous zero-copy injection of mCBM data into the data lake. Registering a replica and a corresponding replication rule took 30-60s per each 4-GB file (having subtracted the aforementioned forced delay), i.e. comparable to the rate at which the data has been replayed

Current bottleneck: calculation of Adler32 checksums (which obviously has to be done at least once, although with a bit of care one can avoid repeating the calculations for unchanged data) on the client side prior to registration of new replicas. This would ideally use checksums calculated by the underlying file system at the time of data being stored (if available and possible to be extracted at file rather than inode level), however Adler32 appears to be supported by rather few modern file systems and may not be available even when supported (e.g. at GSI - Lustre does support Adler32 but our cluster uses CRC-32C for performance reasons)

Possible future development in Rucio: add support for more modern checksum algorithms (e.g. xxhash as fast hash and SHA256/BLAKE2B as strong hash), ideally featuring a transition path from the current Adler32+MD5 schema

## CBM simulation

The future CBM experiment at FAIR in its electron configuration with a MVD, STS, RICH, TRD, TOF and PSD detectors was simulated. (See Fig. 3) This constitutes the most detector rich configuration of the full CBM experiment.

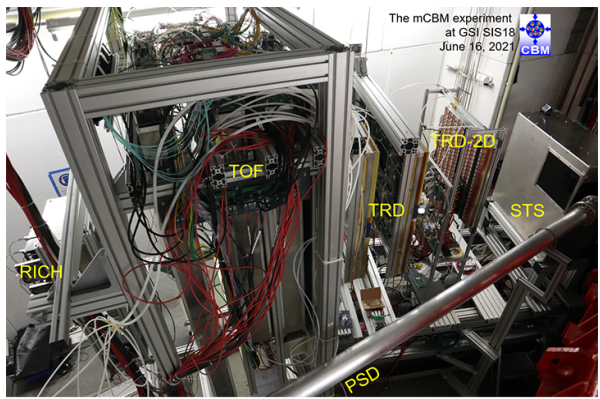

Figure 2: Above image shows the mCBM experimental setup. The beam pipe is visible transversing the lower right quadrant and the target box on the right centre of the photo. Each of the six detectors subsystems are over-labelled.

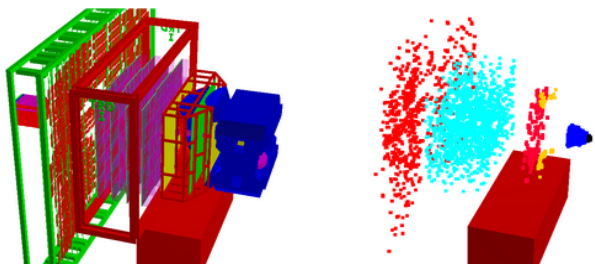

Figure 3: Left shows the CBM simulation geometries for the future CBM experiment at FAIR. It consists of a blue magnet yoke on the right, followed by the RICH detector, the TRD, the TOF and PSD detector on the left. The beam enters from the right. The STS detector is contained inside the magnet. Right shows hits after transport from the demo scripts used during DAC21. Black (Blue) points show hits in the sensors of the MVD (STS) detector.

Task completed during (Thu 2021-11-25) and (Fri 2021-11-26). In the weeks before the challenge, several docker images and demo scripts were prepared. CBMROOT, our simulation software developed at GSI/FAIR for the simulation of the CBM experiments is built upon FairRoot which in turn is built upon FairSoft software packages. Docker images for each of these steps, starting from the most recent stable docker-hub Debian release were produced. Finally a fourth docker image containing the necessary python, gfal, rucio [5], java, voms-client [6] and xrootd [7] was subsequently built on top. This allowed the task to be completed using a standard MacBook connected to a network external to GSI running a docker container.

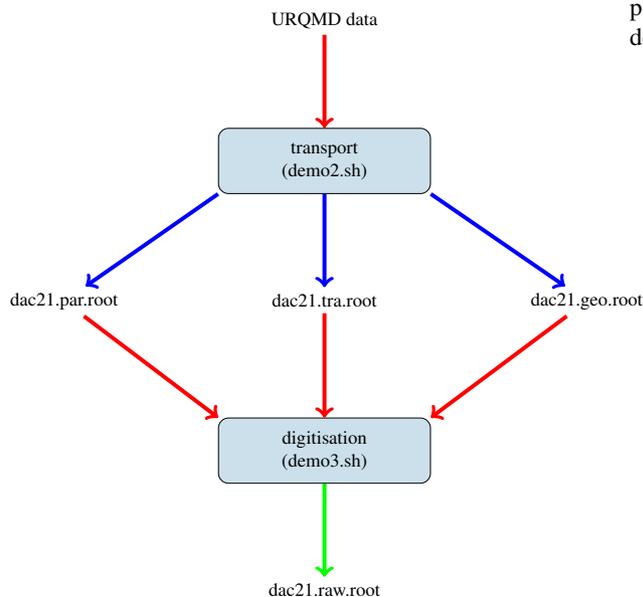

Figure 4: Demos used in simulation challenge. The red arrows show data extracted from data lake, and the blue arrows show data sent to the datalake.

Demo script one (demo1.sh) uploaded initial URQMD data to the data lake using Rucio and followed an extraction from the data lake. The file was consistent with initial interaction of a gold beam of 10AGeV on a gold target. File consistency was verified via md5sum before and after uploads. Subsequent running of script would have returned failure as this data had already been placed on the datalake and therefore would not have been needed to be run during DAC21.

A second demo script (demo2.sh) simulates the transport of particles through the CBM experiment in its electron setup as shown in the left panel of Fig. 3. Three data files are uploaded to the data lake containing parameters, geometry and transport data as symbolically represented by the blue arrows in Fig. 4. It was decided that the demo script would run as a cron job running every 5 minutes which started on Thursday evening and continued through to Friday midnight.

A third script (demo3.sh) extracted some of this transport data and generated digitisation data which was run ad-hocly during the two days. Hits in the sensors of the CBM detector for a typical run is displayed in the right panel of Fig. 3. Demonstration was sporadically successful. It is noted however that interaction with the datalakes gave error messages due to server side issues. Additionally, timeout of voms proxy was also an issue, as some long timelapses between connections occurred. Success and failure statistics are discussed in the proceeding study.

### mCBM reconstruction

Goal of this task was to reconstruct recent mCBM (July 2021) data interacting with the data lake. Task was completed inside a docker container run on a Gentoo Linux desktop machine on the internal GSI network.

- 2021-11-22 It was noted that much of our demo environment developed for this task suddenly became obsolete due to changes in the development branch of CBMROOT which required updated versions of FairSoft and FairRoot. To make matters worse, these were major rather than incremental changes which required modification of the installation process. Bleeding-edge CbmRoot was needed for reconstruction of July data as tracking and reconstruction code were being actively developed. The necessary CBMROOT software was not available to complete this task on this day.

- 2021-11-23 Development of new Docker Images built on latest stable Debian, building a FairSoft image, on which a FairRoot image was built followed by a CbmRoot image. Finally an image with Rucio, gfal libraries, xrootd and all necessary additions to CBMROOT necessary to complete this task was built on top of the CbmRoot image.

- 2021-11-24 Although unpacking of mCBM had been working for a week prior to DAC21, unpackers had

made it into CbmRoot the week prior, several developments in the reconstruction data still had not been submitted to the CBMROOT development branch. It is therefore difficult to prepare for this DAC21 task ahead of time. On this day, our CBMROOT software was still not able to reconstruct the 2021 data and we seriously planned to switch to plan B which meant reconstructing 2020 data instead.

- 2021-11-25 New tracking and reconstruction code merged to the development branch of CbmRoot by its tracking team. These new CBMROOT macros allow reconstruction of the real experiment 2021 data from the SIS18 experiment shown in Fig. 2. The simulation geometries shown in Fig. 5 was used during the reconstruction process.

- 2021-11-26 13:00 CbmRoot source code updated and rebuilt within the running Docker container. This was also done so as to not have to rerun the docker image which was being used for the task.

- 2021-11-26 15:00-19:00 Several demonstration scripts tested and completed our task although in an ad-hoc fashion. Decided a systematic approach is warranted.

- 2021-11-26 18:47 Some raw mCBM data from run 1588 taken during July 2021 is uploaded to the data lake using a Rucio-upload command. This forms the basis for later reconstruction.

- 2021-11-26 20:05 Script (demo6.sh) pulls this mCBM data. If pull is successful then continues, otherwise failure 1 is output to the log file. Next the script reconstructs the data using standard mCBM reconstruction macros merged into CBMROOT the day before. Upload reconstructed data to data lake otherwise declare failure 2 to the log files. Only full completion is considered a success.

- 2021-11-26 20:05 In order to get some success/failure statistics, this demonstration was run one hundred times with the script (demo6.sh) in a "for" loop with a 60 second sleep between runs.

Statistics trial started at 21:04:50 and ended at 23:25:50. Of the 100 cases, 0 had "failure 1", i.e. reading the raw mCBM data from the data lake using a Rucio-get command, 75 had "failure 2", i.e. writing the reconstructed mCBM data to the data lake using a Rucio-upload command, and 25 were deemed fully successful having no failures reported. The failure rates were high, as one of the protocols failed on the clients side. The target RSE supported both "root://" and "davs://" as protocols. When contacting the RSE via the Rucio upload command, the Rucio client primarily chose "davs://" as the protocol. The "davs://" protocol was not supported in the container due to a minor error in the configured paths. gfal2, using its HTTPS plug-in, couldn't resolve the libdavix dependencies correctly. It therefore reported that the protocol is not supported. The error message also contained "The requested service is not available at the moment", which is misleading and lets a user believe that the service was at fault. Despite the high failure rates, overall the challenge was deemed successful as a showcase of reconstruction of mCBM data interacting with the ESCAPE data environment.

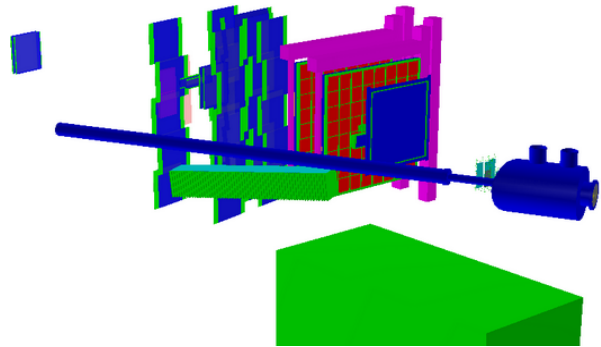

Figure 5: Shows the simulation geometries of the July mCBM setup. The figure is comparable to the photo shown in Fig. 2